\documentclass[fp,onecolumn]{jpsj3}
\usepackage[pdftex]{graphicx} 
\usepackage{amsmath}
\usepackage{bm}
\usepackage[nooneline]{subfigure}
\subfiguretopcaptrue
\usepackage{multirow}
\usepackage{autobreak}
\usepackage{cite}

\title{Bayesian Inference on Hamiltonian Selections for Mössbauer Spectroscopy}
\author{Ryota Moriguchi$^{1}$, Satoshi Tsutsui$^{2,3}$, Shun Katakami$^4$, Kenji Nagata$^5$, Masaichiro Mizumaki$^2$, and Masato Okada$^4$}
\date{\today}
\inst{$^{1}$Graduate School of Science, The University of Tokyo, Bunkyo, Tokyo 113-0033, Japan\\
$^2$Japan Synchrotron Radiation Research Institute (JASRI), Sayo, Hyogo 679-5198, Japan \\
$^3$Institute of Quantum Beam Science, Graduate School of Science and Engineering, Ibaraki University,
Hitachi, Ibaraki 316-8511, Japan \\
$^4$Graduate School of Frontier Sciences, The University of Tokyo, Kashiwa, Chiba 277-8561, Japan\\
$^5$Research and Services Division of Materials Data and Integrated System, National Institute for Materials Science,
Tsukuba, Ibaraki 305-0044, Japan}
\abst{
 Mössbauer spectroscopy, which provides knowledge related to electronic states in materials, has been applied to various fields such as condensed matter physics and material sciences. 
In conventional spectral analyses based on least-square fitting, hyperfine interactions in materials have been determined from the shape of observed spectra. 
In conventional spectral analyses, it is difficult to discuss the validity of the hyperfine interactions and the estimated values. 
We propose a spectral analysis method based on Bayesian inference for the selection of hyperfine interactions and the estimation of Mössbauer parameters. 
An appropriate Hamiltonian has been selected by comparing Bayesian free energy among possible Hamiltonians. 
We have estimated the Mössbauer parameters and evaluated their estimated values by calculating the posterior distribution of each Mössbauer parameter with confidence intervals. 
We have also discussed the accuracy of the spectral analyses to elucidate the noise intensity dependence of numerical experiments. 
}

\begin{document}
\maketitle
\section{Introduction}

Hyperfine interactions, which are interactions between nuclear and electronic systems, provide information on microscopic electronic states and nuclear structures obtained by hyperfine spectroscopies. 
Since a hyperfine structure is described as the product of the parameters of electronic and nuclear systems, nuclear parameters are treated as known constants in the application of hyperfine spectroscopies to material sciences in order to elucidate microscopic electronic states. 
Hyperfine spectroscopies such as nuclear magnetic (quadrupole) resonance [NMR (NQR)] and perturbed angular correlation (PAC) spectroscopies are tools to elucidate microscopic electronic states. 
Since Mössbauer spectroscopy is a nuclear absorption spectroscopy using the nuclear transitions between the excited and ground states, it is usefull for elucidating physical and chemical properties that are difficult to obtain by other hyperfine spectroscopies. 
The characteristic Mössbauer parameters are the recoil-free fraction and isomer shift. 
The former corresponds to the spectral intensity reflecting the probability of the Mössbauer effect coupled with the phonon spectrum at the probe nuclear sites. 
The latter corresponds to the shift of the entire Mössbauer spectrum reflecting the valence and spin states at the probe nuclear sites, when the difference in nuclear radius between the excited and ground states is sufficiently large. 
Nuclear quadrupole interactions reflecting the electron field gradient at the probe nuclear sites and nuclear Zeeman interactions reflecting magnetism, which are observable by NMR (NQR) and PAC spectroscopies as mentioned above, are also observable by Mössbauer spectroscopy. 

Since the observation of the Mössbauer effect in the \(^{191}\mathrm{Ir}\) nucleus \cite{mossbauer1958nuclear}, Mössbauer spectroscopy has been applied to about 40 elements and 80 nuclei as a probe in various fields, mainly in material sciences.\cite{nn1971mossbauer} 
Among these, many applications of \(^{57}\mathrm{Fe}\) Mössbauer spectroscopy have been reported for about two-thirds of the publications related to applications of Mössbauer spectroscopy.\cite{40022493358} 
Although \(^{57}\mathrm{Fe}\) Mössbauer spectroscopy is mainly applied in materials sciences, it is also applied in mineralogical fields, including earth and planetary sciences \cite{maeda1979mossbauer,maeda1981mossbauer,loew1980calculated,frankel1983fe3o4}, and in life sciences such as in studies of hemoglobin in blood and the bioaccumulation of Fe in microorganisms \cite{klingelhoefer2005mossbauer,vandenberghe2013application}. 
Studies in material sciences, such as those on strongly correlated electron systems and complex chemistry, the valence, magnetism, and electronic configuration at nuclear probe sites have been evaluated by analyzing hyperfine interactions at \(^{57}\mathrm{Fe}\) nuclei. 

Mössbauer spectroscopy is applicable to relatively small amounts of samples. 
However, the observed spectra do not always have sufficient precision for spectral analyses, because the thickness of samples is sometimes much smaller than the attenuation length for the \(\gamma\)-ray energy of Mössbauer transition, or the amounts of samples are too small or insufficient for the production of a precipitate or a brand-new material owing to their rarity. 
In addition, the intensity of radio-active sources allowed in laboratories could be so limited that statistics of measured spectra are insufficient. 
Furthermore, the time for measuring spectra is limited when accelerators for proton beams or reactors for neutron beams require the preparation of short-lived sources or synchrotron radiation facilities require the excitation of probe nuclei. 
In these cases, spectral measurements with sufficient statistics might not be possible, and thus, visual spectral analyses are opted for. 

Each component of Mössbauer spectra exhibits hyperfine interactions, which reflect a local environment, at the probe nuclear site as mentioned above. 
The number of components demonstrates the number of sites occupied by the probe nuclei. 
The parameters that determine the spectral shape of each component can be described by a Hamiltonian of hyperfine interactions. 
The hyperfine interactions in Mössbauer spectroscopy are of three types, isomer shift, nuclear quadruple interaction, and nuclear Zeeman interaction: the isomer shift is observed as an overall shift of each component; the nuclear quadrupole interaction and the nuclear Zeeman interaction are observed to determine the spectral shape.  

The Hamiltonian of hyperfine interactions in each spectral component is inferred from the observed spectra. 
Consequently, parameters are estimated by the least-squares method, fitting the sum of the spectra that can be calculated theoretically at each site to the measured data.
This conventional method has the disadvantage that the accuracy of the estimated parameters cannot be determined because it is a point estimate. 
Additionally, the method is trapped in a local optimum solution owing to the initial value dependence during fitting, making it impossible to find a globally optimal solution.
To overcome this disadvantage, we apply the framework developed by Nagata and coworkers\cite{nagata2012,tokuda2017} to the analysis of multipeak spectra using Bayesian estimation to the parameter estimation of each interaction, the so-called ``Bayesian spectroscopy'' \cite{Akai_2018}.
The method based on Bayesian estimation allows one to obtain a posterior probability distribution that includes both the value of each parameter and the accuracy of the estimate.
The use of a priori information, such as experimental conditions and material properties, has also been shown to be advantageous over conventional methods.
We have also reported on the model selection of Hamiltonians in X-ray photoelectron spectroscopy spectra generated using different models of Hamiltonians\cite{mototake2019bayesian}.

In this paper, we report the analyses of Mössbauer spectra with Hamiltonian selection based on Bayesian inference. 
We analyze two typical spectra using generated models. 
The posterior distribution of the selected Hamiltonian is used to estimate the parameter values of the interaction and their accuracy.
The confidence interval of the estimation is evaluated by varying the noise intensity, which shows that the measurement accuracy required for accurate parameter estimation can be evaluated before the measurement.

\section{Formulation}\label{forward_chap}
\subsection{Generative model and recognition model}
In this section, we describe the model for generating a Mössbauer spectrum based on the formulae given in Ref. \citen{voyer2006complete}.
The hyperfine interactions in Mössbauer spectroscopy consist of the isomer shift \(H_{\mathrm{center}}\), nuclear quadrupole interaction \(H_{QI}\) and nuclear Zeeman interaction \(H_{MI}\) in the nuclear states of \(|I,M\rangle\), where \(H_{\mathrm{center}}\) can be represented without the nuclear states. 
The matrix elements of hyperfine interactions at \(^{57}\mathrm{Fe}\) nuclei are written as

\begin{align}
	\begin{aligned}	
	H_{M3/2}&=\left(\begin{array}{cccc}
	  \frac{3}{2} \alpha \cos \theta & -\frac{\sqrt{3}}{2} \alpha \sin \theta e^{i \phi} & 0& 0 \\
	  -\frac{\sqrt{3}}{2} \alpha \sin \theta e^{-i \phi} & \frac{\alpha}{2} \cos \theta & -\alpha \sin \theta e^{i \phi} & 0\\
	  0& -\alpha \sin \theta e^{-i \phi} & -\frac{\alpha}{2} \cos \theta& -\frac{\sqrt{3}}{2} \alpha \sin \theta e^{i \phi} \\
	  0 & 0& -\frac{\sqrt{3}}{2} \alpha \sin \theta e^{-i \phi} & -\frac{3}{2} \alpha \cos \theta
	  \end{array}\right),\\
	H_{Q3/2}&=\left(\begin{array}{cccc}
		3 A &0& \sqrt{3} A \eta & 0 \\
		0 & -3 A & 0 & \sqrt{3} A \eta \\
		\sqrt{3} A \eta & 0& -3 A& 0 \\
		0 & \sqrt{3} A \eta &0& 3 A
		\end{array}\right),  \\
	H_{M1 / 2}&=\left(\begin{array}{cc}
	  -\frac{\beta}{2} \cos \theta & -\frac{\beta}{2} \sin \theta e^{-i \phi} \\
	  -\frac{\beta}{2} \sin \theta e^{+i \phi} & +\frac{\beta}{2} \cos \theta
	  \end{array}\right),\\
	H_c&=E_{\mathrm{center}},
	\label{hamiltonian_eq}
\end{aligned}
\end{align}
where \(\alpha=g_{3/2} \mu_{N} B_{h f} ,\beta=g_{1 / 2} \mu_{N} B_{h f}\) is defined as the nuclear Zeeman interaction, 
\(g_{M}\) is the nuclear \(g\) factor in nuclear spin \(M\),
\(\mu_N\) is the nuclear Bohr magneton, and \(B_{h f}\) is the magnitude of the internal magnetic field at the probe nuclei.
In the present experiment, we set \(B_{h f}\ge 0\).
\(A\) and \(\eta\) are respectively defined as
\begin{align}
	\label{A_equation}
	\begin{aligned}		
	A&=\frac{e Q V_{z z}}{4 I(2 I-1)},\\
	\eta&=\frac{V_{xx}-V_{yy}}{V_{zz}},
    \end{aligned}
\end{align}
where \(e\), \(Q\), and \(V_{kk}\ (k \in \{x, y, z\})\) are the proton charge, the nuclear quadrupole moment in the first excited state of \(^{57}\mathrm{Fe}\) nuclei, 
and the electric field gradient at the probe nuclei when \(z\) is the principal axis respectively.
The basis of the matrix is the spin quantum number state. 
\(H_{M3/2}\) and \( H_{M1/2}\) represent the nuclear Zeeman interaction, \(H_{Q3/2}\) represents the nuclear quadrupole interaction, and \(H_c\) represents the isomer shift.

Thus, when all interactions are valid, the Hamiltonian excluding that of the isomer shift can be written as
\begin{align}
	\label{hamiltonian-one-site-3/2}
	H_{3/2} &= H_{M3/2}+H_{Q3/2},\\	
	H_{1/2} &= H_{M1/2}.
	\label{hamiltonian-one-site-1/2}
\end{align}
In reality, calculations based on the models represented as a Hamiltonian consisting of appropriate hyperfine interactions are required because effective hyperfine interactions depend on the material. 
Using the above Hamiltonian, we can express the energy eigenvalues of the excited and ground states as
\begin{align}
	H_{3/2}\left|e_{j}\right\rangle &= E_{g,j}\left|e_{j}\right\rangle,\\	
	H_{1/2}\left|g_{i}\right\rangle &= E_{e,i}\left|g_{i}\right\rangle.
\end{align}
The transition energy of the absorption process in Mössbauer spectroscopy is calculated from the difference between eigenenergies.
However, since \(E_{\mathrm{center}}\) is independent of the magnetic quantum number of nuclear spins, we obtain the following as
\begin{align}
	E_{i,j} = E_{e,i} - E_{g,j}+E_{\mathrm{center}}.
	\label{energy_dif}
\end{align}
Next, the absorption intensity is calculated.
When the eigenstates of nuclear spin \(I\) are expanded in terms of the spin quantum state, the coefficients can be written as follows,
\begin{align}
	\left|g_{i}\right\rangle
	=\left(\begin{array}{c}
		\left\langle m_{g_{i}}=I \mid n_{g_{i}}\right\rangle \\
		\left\langle m_{g_{i}}=I-1 \mid n_{g_{i}}\right\rangle \\
		\left\langle m_{g_{i}}=I-2 \mid n_{g_{i}}\right\rangle \\
		\vdots
	\end{array}\right),\quad
	\left|e_{j}\right\rangle
	=\left(\begin{array}{c}
		\left\langle m_{e_{j}}=I \mid n_{e_{j}}\right\rangle \\
		\left\langle m_{e_{j}}=I-1 \mid n_{e_{j}}\right\rangle \\
		\left\langle m_{e_{j}}=I-2 \mid n_{e_{j}}\right\rangle \\
		\vdots
	\end{array}\right).	
	\label{expansion_coefficient}	
\end{align}
Using these expansion coefficients, we calculate the transition probability to each level as
\begin{align}
	a_{i,j}(\mathbf{L}, m)=\sum_{m_{e_{j}}-m_{g_{i}}=m}\left\langle m_{e_{j}} \mid n_{e_{j}}\right\rangle\left\langle m_{g_{i}} \mid n_{g_{i}}\right\rangle^{*}\left\langle I_{g_{i}} \mathbf{L} m_{g_{i}} m \mid I_{e_{j}} m_{e_{j}}\right\rangle,
	\label{transition_probability}
\end{align}
where \(a_{i,j}(\mathbf{L}, m)\) is the transition probability from the ground state energy level \(E_{g,j}\) to \(E_{e,i}\),
and \(\left\langle I_{g_{i}} \mathbf{L} m_{g_{i}} m \mid I_{e_{j}} m_{e_{j}} m_{e_{j}}\right\rangle\) is the Clebsch–Gordan coefficient defined as
\begin{align}
	\left\langle I_{g_{i}} \mathbf{L} m_{g_{i}} m \mid I_{e_{j}} m_{e_{j}}\right\rangle=(-1)^{I_{g_{i}}-m_{g_{i}}} \sqrt{\frac{2 I_{e_{j}}+1}{2\left(I_{e_{j}}-I_{g_{i}}\right)+1}}\left\langle I_{e_{j}} I_{g_{i}} m_{e_{j}}-m_{g_{i}} \mid \mathbf{L} m\right\rangle.
\end{align}
The intensity of the \(^{57}\mathrm{Fe}\) Mössbauer spectrum, that is, the M1 transition from \(I =1/2\) to \(I = 3/2\),
can be written as follows using Eq. \eqref{transition_probability}, which is the transition probability for powder samples:
\begin{align}
	I_{i,j} =|a_{i,j}(1,1)|^{2}+|a_{i,j}(1,0)|^{2}+|a_{i,j}(1,-1)|^{2}.
		 \label{intensity}
\end{align}
From Eqs. \eqref{energy_dif} and \eqref{intensity}, the physical model of Mössbauer spectra can be described by the following Lorentzian:
\begin{align}
	f(x;\Theta) = \sum_{i,j}\frac{1}{\pi}\frac{I_{i,j}\times\Gamma}{(x-E_{i,j}-E_{\mathrm{shift}})^2+\Gamma^2},
	\label{one_site_model}
\end{align}
where the parameter set \(\Theta=\{A,B_{hf},\eta,\theta,\phi,E_{\mathrm{center}},\Gamma\}\) represents the parameter set required for the physical model.
However, in this study, experiments were conducted only for \(E_{\mathrm{shift}}=0\), so \(E_{\mathrm{shift}}\) is not included in the parameter set.

In addition, the generative model is one that adds Gaussian noise to Eq. \eqref{one_site_model} as
\begin{align}
	y &= f(x;\Theta)+\varepsilon,\\
	\varepsilon  &\sim \mathcal{N}(0,\sigma_{\mathrm{noise}}),
\end{align}
where the Gaussian noise is the center \(0\), the standard deviation is \(\sigma_{\mathrm{noise}}\), \(x\) is the channel, and \(y\) is the relative transmission.
The two models generated by this process are shown in Fig. \ref{rowdata_onesite}.
Figure \ref{rowdata_cQ} shows the spectrum of a Hamiltonian that considers only quadrupole splitting \(H_c+H_Q\) and \(\sigma_{\mathrm{model}}=0.01\).
Figure \ref{rowdata_cQM} shows the spectrum of the Hamiltonian \(H_c+H_Q+H_M\), where the magnetic and quadrupole splittings are taken into account and \(\sigma_{\mathrm{model}}=0.003\).
The parameters used in the calculations are shown in Table \ref{table_one_site}.

In this study, the same physical model as the generative model is used as the recognition model for the Mössbauer spectrum as follows:
\begin{align}
	y &= f(x;\Theta)+\varepsilon,\\
	\varepsilon  &\sim \mathcal{N}(0,\sigma_{\mathrm{model}}).
\end{align}
By this analysis method, we can estimate the parameter \(\Theta\) in the recognition model such that it reproduces the generative model.
\subsection{Posterior distribution}
Here, we describe the posterior distribution \cite{nagata2012} in Bayesian inference.
In the aforementioned recognition model, the probability of obtaining the measurement \(y\) for a given \(x\) is calculated as
\begin{align}
p(y|x,\Theta)=\frac{1}{\sqrt{2\pi \sigma_{\mathrm{model}}^2}}\exp \left(-\frac{1}{2\sigma_{\mathrm{model}}^2}(y-f(x;\Theta))^2\right).
\end{align}
Assuming that the observed data \(D=\{y_i,x_i\}(i=1,\dots,n)\) are obtained independently, the likelihood function is
\begin{align}
p(D|\Theta)=\prod^n_{i=1}p(y_i|x_i,\Theta)\propto \exp \left(-\frac{n}{\sigma_{\mathrm{model}}^2}E_n(\Theta)\right),
\label{likelyfood}
\end{align}
where the function \(E_n(\Theta)\) is defined as
\begin{align}
  E_n(\Theta)=\sum^n_{i=1} \frac{1}{2n}(y_i-f(x_i;\Theta))^2.
\end{align}
From the likelihood function [Eq. \eqref{likelyfood}], the posterior distribution \(p(\Theta|D)\) can be calculated using Bayes' theorem as
\begin{align}\label{posterior}
  p(\Theta|D)&=\frac{p(D|\Theta)p(\Theta)}{p(D)}\nonumber,\\
   &= \frac{1}{Z(D)}\exp\left(-\frac{n}{\sigma_{\mathrm{model}}^2}E_n(\Theta)\right)\varphi(\Theta).
\end{align}
In this study, we define the prior distribution \(\varphi(\Theta)\) as
\begin{align}
	\varphi(\Theta)=\varphi_A(A)\varphi_{B_{hf}}(B_{hf})&\varphi_\eta(\eta)\varphi_{\theta}(\theta)\varphi_{\phi}(\phi)\varphi_{E_{\mathrm{Center}}}(E_{\mathrm{Center}})\varphi_{\Gamma}(\Gamma),\\
	\varphi_{\Gamma}(\Gamma) &=\operatorname{Gamma}\left(\Gamma ; \eta, \lambda\right), \nonumber\\
	&=\frac{1}{\operatorname{G}\left(\eta\right)}\left(\lambda\right)^{\eta}\left(\Gamma\right)^{\eta-1} \exp \left(-\lambda \Gamma\right),\label{gamma_prior}\\
	\varphi_x(x) &=\operatorname{Uniform}\left(x ; x_{max}, x_{min}\right), \\
	&x=(A,B_{hf},\eta,\theta,\phi,E_{\mathrm{Center}}),
\end{align}
where \(\operatorname{Uniform}\left(x ; x_{max}, x_{min}\right)\) is the uniform distribution of the maximum value \(x_{max}\) and minimum value \(x_{min}\), and \(\operatorname{G}(\cdot)\) is the gamma function.

In this study, only the parameter \(\Gamma\), which is related to spectral width, is used as the gamma distribution, as in Eq. \eqref{gamma_prior}.
In Mössbauer spectroscopy, the spectral width inequality can be obtained from the energy uncertainty relation
\begin{align}
	\tau \cdot \Gamma \geq h,
	\label{uncertainty_principle}
\end{align}
where \(\tau\) is the lifetime of the excited state in the Mössbauer transition of probe nuclei, \(\Gamma\) is the spectral width, and \(h\) is Planck's constant.
From Eq. \eqref{uncertainty_principle}, the spectral width is not smaller than the natural width, so the prior distribution of the width is assumed to be a gamma distribution.

\subsection{Exchange Monte Carlo method}\label{ex_chap}
In this study, the posterior distribution of the parameters is obtained by the exchange Monte Carlo (EMC) sampling method\cite{hukushima1996EMC}.

The EMC method involves sampling by exchanging samples among multiple replicas as one replica of a probability distribution with an inverse temperature \(\beta\) such as that in Eq. \eqref{EMC_replica_eq}.
\begin{align}
	p(\Theta|D,\beta)&= \frac{1}{Z(D)}\exp\left(-\frac{n}{\sigma_{\mathrm{model}}^2}\beta E_n(\Theta)\right)\varphi(\Theta).
	\label{EMC_replica_eq}
\end{align}
In this study, each replica layer was sampled by the Metropolis method\cite{metropolis1953equation}.
The simultaneous posterior probabilities for all these replicas can be written as
\begin{align}
	p(\Theta|D)&=\prod^L_{l=1} p(\Theta_l|D,\beta_l), 
	\label{EMC_posterior}
\end{align}
where \(\Theta_l\) represents the physical model parameters for the \(l\)th replica layer.
The inverse temperature \(\beta \in \{\beta_1,\dots,\beta_{l+1}\}\) is set to \(\beta_l < \beta_{l+1}\) for any \(l\).

\subsection{Estimation and confidence intervals}
In our method, we use the following maximum a posteriori (MAP) estimate \(\Theta_{\mathrm{map}}\), the parameter with the maximum posterior probability, to estimate the parameters:
\begin{align}
	\Theta_{\mathrm{map}} = \underset{\Theta}{\mathrm{argmax}}[p(\Theta|D,\beta=\beta_{ac})],
\end{align}
where \(\beta_{ac}\) is set so that \(\sigma_{\mathrm{model}}/\sqrt{\beta_{ac}}\) is close to the standard deviation of Gaussian noise in the measured \(\sigma_{\mathrm{noise}}\) values.
In this study, the estimated Bayesian free energy noise is used to determine \(\beta_{ac}\).
The standard deviation of the sample is calculated as the spread of the posterior distribution and used as a confidence interval for the MAP estimate to evaluate the estimation accuracy.

\subsection{Bayesian free energy for Hamiltonian selection}
Hamiltonian and noise intensity are estimated using Bayesian free energy\cite{tokuda2017}.
In this section, we explain how to calculate the Bayesian free energy.
By using a sample of the posterior distribution of each replica in the EMC of Sect. \ref{ex_chap}, we calculate the distribution function\cite{Neal93probabilisticinference} \(Z\left(\beta_{m}\right)(1 \leq m \leq L)\) as
\begin{align}
	\tilde{Z}_n(\beta_m), 
	&=\frac{\tilde{Z}_n\left(\beta_{m}\right)}{\tilde{Z}_n\left(\beta_{m-1}\right)} \times \cdots \times \frac{\tilde{Z}_n\left(\beta_{2}\right)}{\tilde{Z}_n\left(\beta_{1}\right)},\\
	&=\prod_{l=1}^{m-1} \int \frac{\tilde{p}(\Theta_l|D,\beta_{l+1})}{\tilde{p}(\Theta_l|D,\beta_{l})}p(\Theta_l|D,\beta_{l})d\Theta_l,\\
	&=\prod_{l=1}^{m-1}\left\langle\exp \left(-\frac{n}{\sigma_{\mathrm{model}}^{2}}\left(\beta_{l+1}-\beta_{l}\right) E_n(\Theta_l)\right)\right\rangle_{p(\Theta_l|D,\beta_{l})},
\end{align}
where \(p(\Theta_l|D,\beta_{l})=\tilde{p}(\Theta_l|D,\beta_{l})/\tilde{Z}_n\left(\beta_{l}\right)=\int \tilde{p}(\Theta_l|D,\beta_{l}) d\Theta_l\).
By numerically obtaining the distribution function using the above method, the Bayesian free energy\cite{tokuda2017} \(F_{n}(\beta)\) as

\begin{align}
	F_{n}(\beta) &:=-\log Z_{n}(\beta), \\
	&=\beta \tilde{F}_{n}(\beta)-\frac{n}{2}(\log \beta-2\log\sigma_{\mathrm{model}}-\log 2 \pi), \\
	\tilde{F}_{n}(\beta) &:=-\frac{1}{\beta} \log \tilde{Z}_{n}(\beta).
\end{align}

The distribution function corresponds to the marginal likelihood, and the Bayesian free energy represents the negative log marginal likelihood.
Therefore, the lower the Bayesian free energy, the better the model.
\section{Numerical Experiment}
\label{experiment}
\subsection{Hamiltonian selection and noise estimation}
In this study, to verify the validity of Bayesian inference for the Mössbauer spectrum, we conducted experiments with spectral data using a generative model of Eq. \eqref{one_site_model}.
The generated models used were the Hamiltonian \(H_c+H_Q\) in the absence of nuclear Zeeman interaction and the Hamiltonian \(H_c+H_Q+H_M\) in the presence of nuclear Zeeman interaction.
Figure \ref{rowdata_onesite} shows the spectra from each of the generative models. 

The recognition models used to analyze the spectra are the four patterns of Hamiltonians \(H_c\), \(H_c+H_Q\), \(H_c+H_M\), and \(H_c+H_Q+H_M\) in Eq. \eqref{one_site_model}.
The EMC method with each Hamiltonian recognition model is performed with a number of replicas \(L=64\), \(200,000\) iteration steps, and \(100,000\) burn-in steps.

In this method, we simultaneously perform noise estimation and Hamiltonian selection by determining the Bayesian free energy for each replica of each recognition model.
We estimate noise by determining the standard deviation of Gaussian noise \(\sigma_{\mathrm{model}}/\sqrt{\beta}\) from the inverse temperature \(\beta\) of the replica with the lowest Bayesian free energy.
In addition, we estimate the Hamiltonian by selecting the Hamiltonian of the recognition model in the selected replicas.
We obtain the posterior distribution numerically by the EMC method using the physical model substituted into Eq. \eqref{EMC_posterior}.
Table \ref{table_prior_params} shows the parameters of the prior distribution.

\begin{table}
	\caption{Parameters of the prior distribution}
	\label{table_prior_params}
	\centering
	 \begin{tabular}{c||c|c|c|c|c|c||c|c}
	  \hline
	    & \(\varphi_A(A)\)&\(\varphi_{B_{hf}}(B_{hf})\)& \(\varphi_{\eta}(\eta)\)&\(\varphi_{\theta}(\theta)\)&\(\varphi_{\phi}(\phi)\)& \(\varphi_{E_{\mathrm{Center}}}(E_{\mathrm{Center}})\)&& \(\varphi_{\Gamma}(\Gamma)\)\\
	  \hline \hline
	  \(x_{max}\)&10.0&100&1.0&\(\pi\)&2\(\pi\)&500.0&\(\eta\)&4.0\\\hline
	  \(x_{min}\)&-10.0&0.0&0.0&0.0&0.0&0.0&\(\lambda\)&1.0\\\hline
	 \end{tabular}
\end{table}
  \begin{table}
	\caption{Parameters of the generative model}
	\label{table_one_site}
	\centering
	 \begin{tabular}{c||c|c|c|c|c|c|c}
	  \hline
	  Hamiltonian  & \(A\)&\(B_{hf}\)& \(\eta\)&\(\theta\)&\(\phi\)& \(E_{{\rm Center}}\)& \(\Gamma\)\\
	  \hline \hline
	  \(H_c+H_Q\)&-3.0&&1.0&&&250.0&3.0\\\hline
	  \(H_c+H_Q+H_M\)&-3.0&40&1.0&0&0&250.0&3.0\\\hline
	 \end{tabular}
   \end{table}

\subsection{Limits of analyzable noise intensity}
In this section, we examine how much noise can be analyzed with our method, that is, we determine at what accuracy of Mössbauer experimental data can the proposed method be applied.
To varify the accuracy, we conducted on experiment to see how the MAP estimate and confidence interval change with the standard deviation of the Gaussian noise in the generated model.
We define the confidence intervals as the standard deviations in the sample of posterior distributions obtained by the EMC method.

We used a physical model as the generative model with the parameters from Table \ref{table_one_site}.
The standard deviations of the added Gaussian noise were \(\sigma_{\mathrm{noise}}=10^{i/2}(i=0,-1,\dots,-6)\), and the posterior distribution of the parameters was sampled for each standard deviation to obtain confidence intervals for the estimated values.
Noise was not estimated here to test how the estimates and confidence intervals change. 
The standard deviation of the Gaussian noise in the posterior distribution of each replica [Eq. \eqref{posterior}] was set to \(\sigma_{\mathrm{model}}=\sigma_{\mathrm{noise}}\) and the inverse temperature to \(\beta_1=0,<\dots,<\beta_L=1\).
At each noise intensity, the EMC method was performed with a number of replicas \(L=32\), \(200,000\) iteration steps, and \(100,000\) burn-in steps.

\section{Results and Discussions}
\label{one_site_result}
\subsection{Hamiltonian selection and noise estimation}
\begin{figure}[htbp]
	\begin{center}	
	\subfigure[\label{rowdata_cQ}]{%
		\includegraphics[clip, width=0.5\columnwidth]{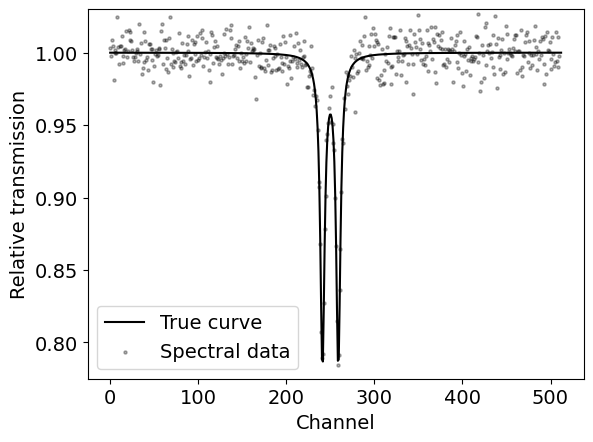}}%
	\subfigure[\label{rowdata_cQM}]{%
		\includegraphics[clip, width=0.5\columnwidth]{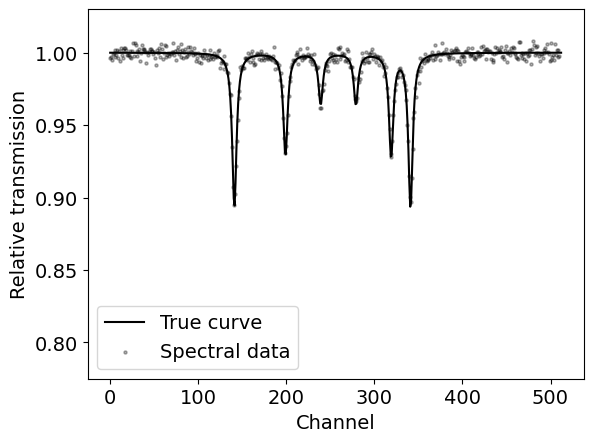}}%
	\caption{
		    Spectral data of the generative model.
			The gray dots are spectral data, and the black lines are the true curves of the data.
			(a) Hamiltonian \(H_c+H_Q\) and Gaussian noise with standard deviation \(\sigma_{\mathrm{noise}}=0.01\).
			(b) Hamiltonian \(H_c+H_Q+H_M\) and Gaussian noise with standard deviation \(\sigma_{\mathrm{noise}}=0.003\).
			\label{rowdata_onesite}} 
	\end{center}
\end{figure}

\begin{figure}[htbp]
	\begin{center}		
	\subfigure[]{\label{f_ene_cQ}%
		\includegraphics[clip, width=0.5\columnwidth]{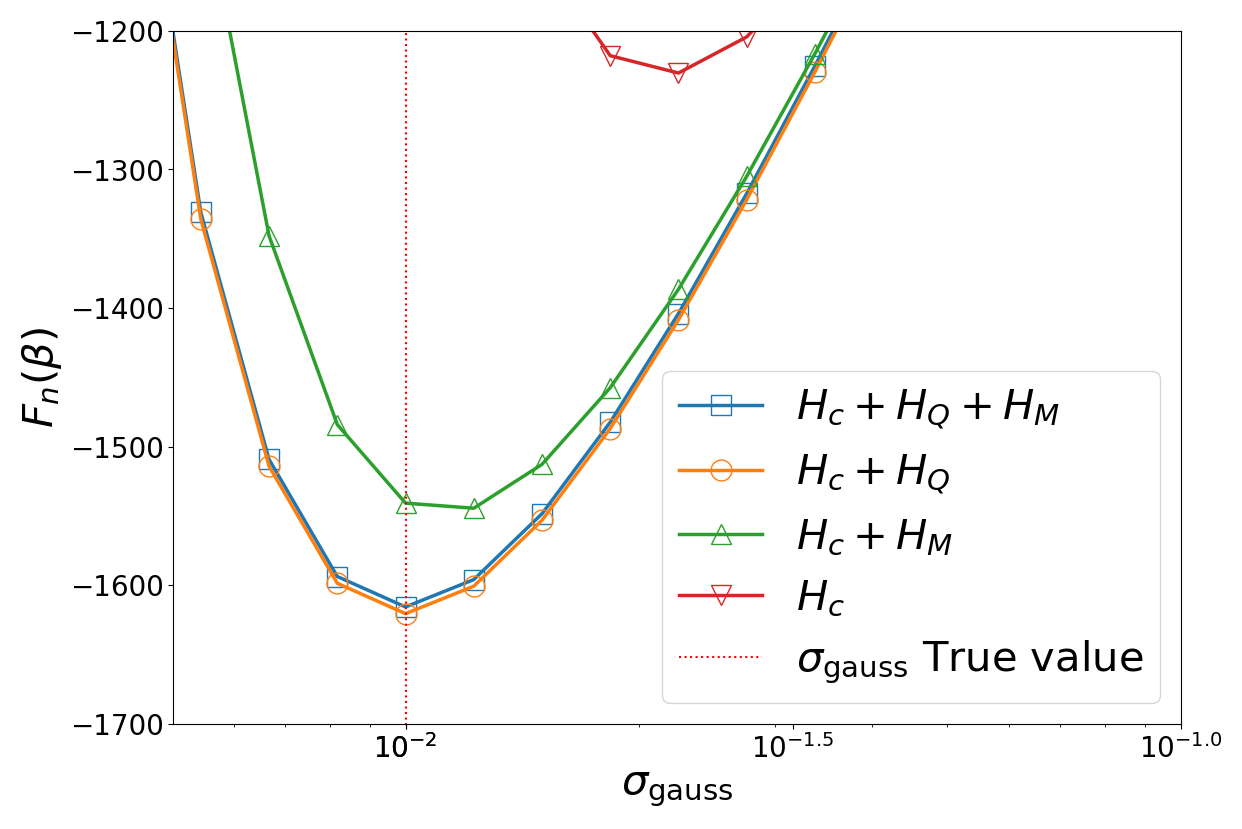}}%
	\subfigure[]{\label{f_ene_cQM}%
		\includegraphics[clip, width=0.485\columnwidth]{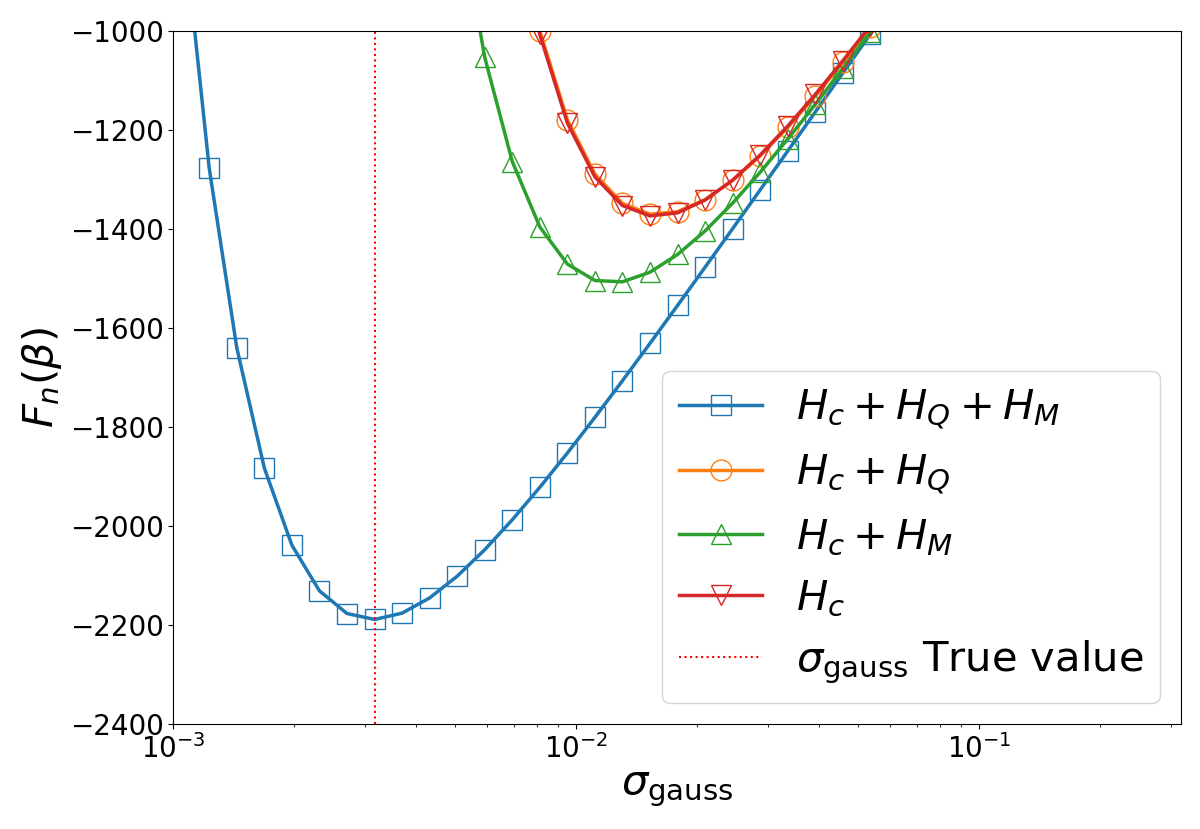}}%
	\caption{
			(Color online) Bayesian free energy for each experiment. 
			The horizontal axis represents the standard deviation of the Gaussian noise of the recognition model \(\sigma_{\mathrm{gauss}}=\sigma_{\mathrm{model}}/\sqrt{\beta}\) on log scale, 
			the vertical axis represents the Bayesian free energy \(F_n(\beta)\) for each recognition model, 
			and the red dotted lines are the true values of the standard deviation.
			(a) Generative model of \(H_c+H_Q\) for standard deviation of 0.01 and (b) generative model of \(H_c+H_Q+H_M\) for standard deviation of 0.003.}
	\label{f_ene_onesite} 
	\end{center}
\end{figure}

\begin{figure}[htbp]
	\begin{center}		
	\subfigure[]{\label{hist_cQ}%
		\includegraphics[clip, width=0.35\columnwidth]{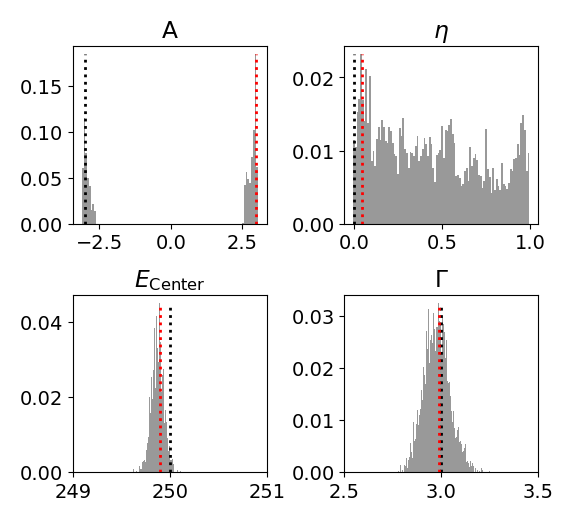}}%
	\subfigure[]{\label{hist_cQM}%
		\includegraphics[clip, width=0.5\columnwidth]{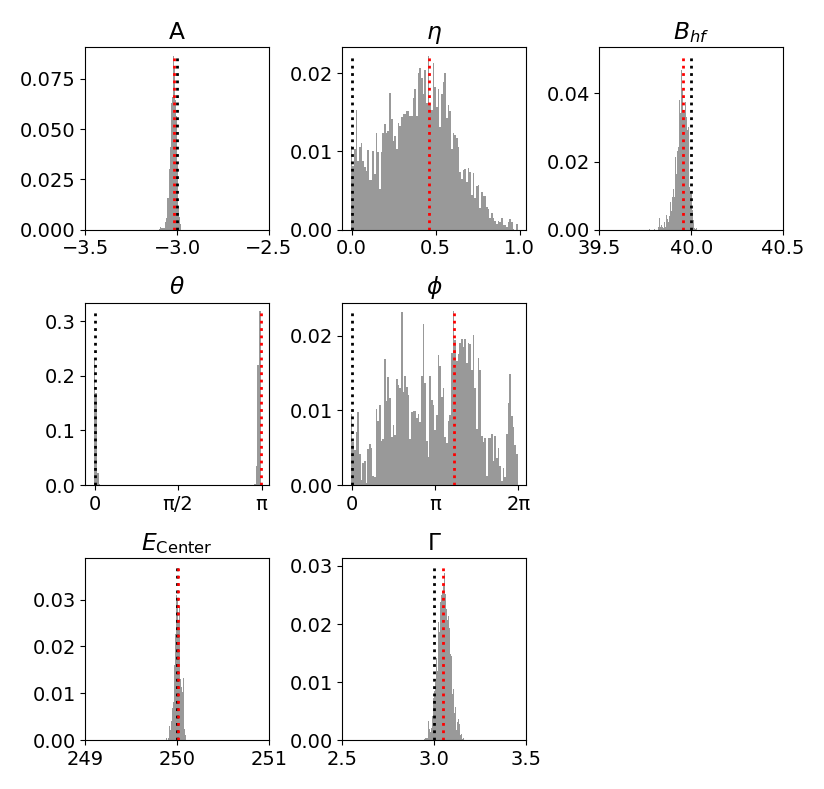}}%
	\caption{
		    (Color online) Histograms of the posterior distributions for Hamiltonians (a) \(H_c+H_Q\) and (b) \(H_c+H_Q+H_M\).
			The red dotted lines are the estimated values, and the black dotted lines are the true values of the parameter.}
	\label{hist_onesite} 
	\end{center}
\end{figure}

\begin{table}
	\caption{Estimated values and confidence intervals for each parameter at the artificial data using parameters in Table \ref{table_one_site}.}
	\label{table_onesite_map}
	\centering
	 \begin{tabular}{c|c||p{1cm}|p{1cm}|p{1cm}|p{1cm}|p{1cm}|p{1cm}|p{1cm}}
	  \hline
	  Hamiltonian &  & \(A\)&\(B_{hf}\)& \(\eta\)&\(\theta\)&\(\phi\)& \(E_{{\rm Center}}\)& \(\Gamma\)\\
	  \hline \hline
	  \multirow{2}{*}{\(H_c+H_Q\)}&Estimated values&\(2.996\)&&\(3.515\times 10^{-2}\)&&&\(249.9\)&\(2.965\)\\ \cline{2-9}
	  &Confidence intervals&\(2.878\)&&\(2.935\times 10^{-1}\)&&&\(8.152\times 10^{-2}\)&\(8.133\times 10^{-2}\)\\\hline\hline
	  \multirow{2}{*}{\(H_c+H_Q+H_M\)}& Estimated values &\(-3.023\)&\(39.97\)&\(5.611\times 10^{-1}\)&\(3.126\)&\(4.619\)&\(250.0\)&\(3.054\)\\ \cline{2-9}
	  &Confidence intervals &\(6.761\times 10^{-1}\)&\(4.321\times 10^{-2}\)&\(2.324\times 10^{-1}\)&&&\(4.074\times 10^{-2}\)&\(3.993\times 10^{-2}\)\\ \hline
	 \end{tabular}
\end{table}
First, for the generative model shown in Fig. \ref{rowdata_cQ}, we analyzed with the method described in the previous section, using the four patterns of the recognition model, \(H_c\), \(H_c+H_Q\), \(H_c+H_M\), and \(H_c+H_Q+H_M\) as selections under the condition that the nucleus can take a single state.
We computed posterior probabilities in the model Hamiltonian for each selection and calculated Bayesian free energy on the basis of the obtained analysis results.
Figure \ref{f_ene_cQ} shows the calculated Bayesian free energy plotted against the inverse temperature \(\beta\) (standard deviation of Gaussian noise). 
Each model Hamiltonian exhibits a minimum for a specific standard deviation of Gaussian noise.
The minima show different values for each model Hamiltonian of selection.
Of the minima of the Bayesian free energy for each model Hamiltonian shown in Fig. \ref{f_ene_cQ}, the model Hamiltonian that gives the smallest value was selected as the best model Hamiltonian to reproduce the data in Fig. \ref{rowdata_cQ}.
As a result, we found that the Hamiltonian described by the sum of \(H_C\) and \(H_Q\), which is the same as the generated model, gives the smallest value, indicating that this method selects an appropriate model Hamiltonian when only nuclear quadrupole interactions occur at the \(^{57}\mathrm{Fe}\) nuclear sites.
Each Mössbauer parameter given by the model Hamiltonian selected from the plots shown in Fig. \ref{f_ene_cQ} gave a posterior probability distribution as in Fig. \ref{hist_cQ}.
For the nuclear quadrupole interaction \(A\), the results show that the posterior probabilities exist at around \(A = +3\) and around \(A = -3\), which is the true value of the parameter given as the artificial spectrum.
The MAP estimate was \(A = 2.996\) which is not the true value of the parameter, \(A = -3\). 
The asymmetric parameter \(\eta\) yielded posterior probabilities that were widely distributed over the range of possible values of \(0\)–\(1\) in principle.
For the isomer shift \(E_{\mathrm{center}}\) and the width at half maximum \(\Gamma\), the true values of the parameters and the MAP estimates are in good agreement.

\begin{figure}[htbp]
	\begin{center}		
	\subfigure[]{\label{fit_cQ} %
		\includegraphics[clip, width=0.5\columnwidth]{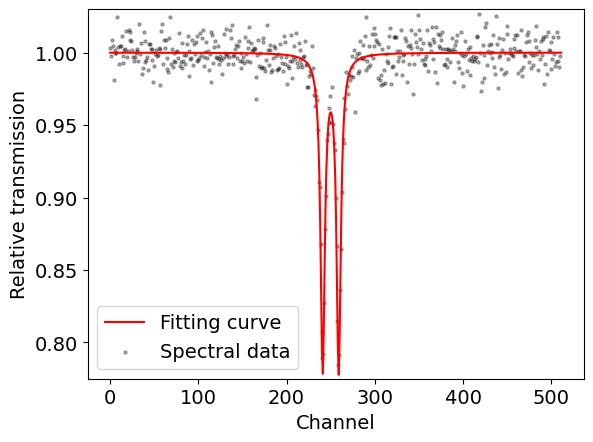}}%
	\subfigure[]{\label{fit_cQM}%
		\includegraphics[clip, width=0.5\columnwidth]{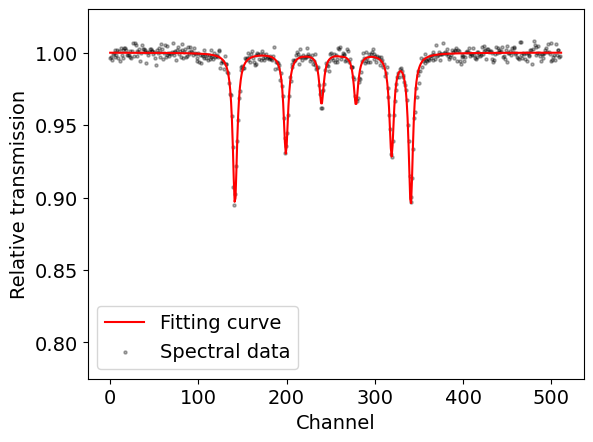}}%
		\caption{
		(Color online) Results of fitting with the correct Hamiltonians, (a) \(H_c+H_Q\) and (b) \(H_c+H_Q+H_M\),
		and the spectral data of the generative model.
		The gray dots are artificial data, and the red lines are the fitting curves.}
	\label{fit_onesite} 
	\end{center}
\end{figure}

Next, for the generative model shown in Fig. \ref{rowdata_cQM}, we performed the same analysis as in Fig. \ref{rowdata_cQ} above with the condition that the nucleus can be in a single state and with the three interaction combinations \(H_C\), \(H_Q\), and \(H_M\) as selections.
We also computed posterior probabilities in the model Hamiltonian for each selection and calculated Bayesian free energies using the analytical results obtained.
Figure \ref{f_ene_cQM} shows the Bayesian free energies obtained after calculating the posterior probabilities for each model Hamiltonian against the inverse temperature \(\beta\).
Among the Bayesian free energy minima for each model Hamiltonian shown in Fig. \ref{f_ene_cQM}, we found that the model Hamiltonian that gives the minimum can be described by the sum of \(H_c\), \(H_Q\), and \(H_M\).
That is, we found that the model Hamiltonian selected by this method is the same as that used in the generative model, even when nuclear Zeeman and nuclear quadrupole interactions coexist.
Each Mössbauer parameter given by the model Hamiltonian selected from the plots shown in Fig. \ref{f_ene_cQM} also gave a posterior probability distribution as in Fig. \ref{hist_cQM}.
Narrow posterior probability distributions at around the true values of the parameters can be shown in the main axis component of the nuclear quadrupole interaction, \(A\), nuclear Zeeman interaction \(B_{hf}\), the angle \(\theta\) between the electric field gradient and the internal magnetic field, the isomer shift \(E_{\mathrm{center}}\), and the line width \(\Gamma\). 
In addition, MAP estimates also show good agreement with the true values of these parameters.
The asymmetric parameter \(\eta\) shows a posterior probability distribution with a maximum at around \(\eta=0.5\) and the distribution over almost the entire possible range of \(\eta\).
The posterior probability distribution of the azimuth angle between the principal axis of the electric field gradient and the internal magnetic field is almost uniform from \(0\) to \(2\pi\).
Figure \ref{fit_onesite} shows the generative model and the regression curves obtained using the MAP estimates from the analysis of the generative models shown in Figs. \ref{rowdata_cQ} and \ref{rowdata_cQM} on the same axis.
We illustrated that the obtained MAP estimates reproduce well the given generative model.
On the other hand, some of the posterior probability distributions given in Fig. \ref{hist_onesite} can not be determined owing to physical requirements.
The reason for this is discussed below.

For some of the parameters shown in Fig. \ref{hist_onesite}, the posterior probability distribution is uniformly distributed over the entire range of possible values of that parameter.
Although the uniform posterior probability distribution of \(\theta\) in Fig. \ref{hist_cQ} affects at least the posterior probability distribution of the magnitude of the nuclear quadrupole interaction \(|A|\), \(\phi\) in Fig. \ref{hist_cQM} is independent of the magnitude of the hyperfine interaction \(A\) or \(B_{hf}\).
The former is a peculiar problem for sublevel eigenvalues, which occurs only in the \(I = 3/2\) state.
As is discussed, the posterior probability distribution indeed exhibits the presence of unresolved problems related to relationships among various Mössbauer parameters. However, when the spectral linewidth of \(\Gamma\) = 3.0, where the posterior probability distribution exhibits a maximum, is \(0.3\ \mathrm{mm/s}\) of the resolution function given by normal Mössbauer spectrometers, the obtained results indicate remarkable success in extracting Mössbauer parameters reasonably. 
In this case, the posterior probability distribution is caused by the correlation among various parameters, and the spectra are not necessarily the same as those obtained with the Mössbauer spectrometer usually used. 
The results show that the parameters can be extracted with high accuracy.
The following is a discussion of the factors contributing to the posterior distribution of the parameters in each spectrum.
\subsubsection{Application of Bayesian inference to a paramagnetic spectrum}
As shown in Fig. \ref{hist_cQ}, the posterior probability distribution also existed in the region where the parameters were not true values.
This is not a problem with this analytical method, which essentially requires prior information in this case.
The sign of the nuclear quadrupole interaction indicates the spread of the charge distribution corresponding to the positive or negative electric field gradient at the nuclear site, which has a different intrinsic meaning: oblate or elongated charge distribution.
However, in the case of a powder sample, the spectrum expected to be observed is an asymmetric doublet.
This makes it impossible to experimentally distinguish which transitions correspond to absorption lines on the high and low Doppler velocity sides regardless of whether or not \(\eta\) takes finite values other than \(\theta\).
For this reason, the posterior probability distribution based on Bayesian estimation is not only at around the true value, \(A = -3\), as shown in Fig. \ref{hist_cQ}, but also at around \(A = +3\).
We can obtain the spectral splitting width given by the nuclear quadrupole interaction by calculating the eigenvalues of each level from Eq. \eqref{hamiltonian-one-site-3/2} as follows:
\begin{align}
	\Delta E=\left|6A\sqrt{1+\frac{\eta^2}{3}}\right|.
	\label{E_split_eq}
\end{align}

\begin{figure}
	\centering
	\includegraphics[keepaspectratio,scale=0.6]{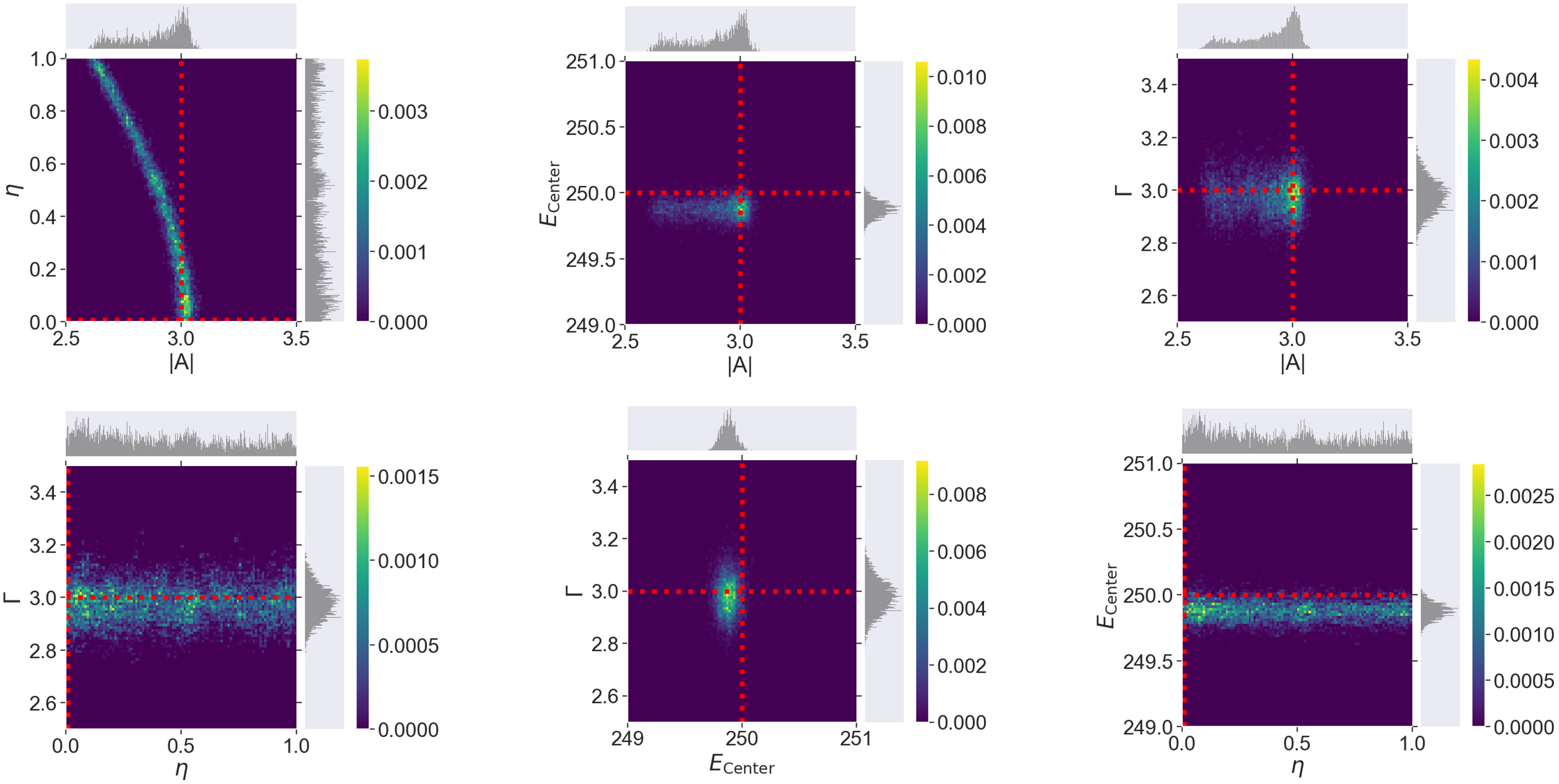}
	\caption{
		(Color online) Two-dimensional histograms of the posterior distributions with the generative model of Hamiltonian \(H_c+H_Q\) and recognition model of the same Hamiltonian.
		The red dotted lines are the true values of each parameter.}
	\label{2dhist_cQ}
\end{figure}

From the above equation, the magnitude of the nuclear quadrupole interaction obtained from Mössbauer experiments is a product of the main axis component of the nuclear quadrupole interaction, \(A\), and a function of the asymmetry parameter \(\eta\).
Therefore, it is impossible to uniquely determine the combination that satisfies \(A\) and \(\eta\) simultaneously.
This is clearly shown in the correlation diagram between the parameters shown in Fig. \ref{2dhist_cQ}.
The steep maxima of the posterior probability distribution are represented by a combination of \(A\) and \(\eta\) that satisfies the relationship shown in Eq. \eqref{E_split_eq} to keep the \(\Delta E\) constant.
In other words, Fig. \ref{2dhist_cQ} shows that if either \(A\) or \(\eta\) can be determined with prior information, then \(A\) and \(\eta\) can be extracted as unique parameters.
The uniform distribution of the posterior probability distribution of \(\eta\) between 0 and 1 that \(\eta\) can take, as in Figs. \ref{hist_cQ} and \ref{2dhist_cQ}, is a unique phenomenon observed only when the nuclear spin is \(I = 3/2\).
When the nuclear spin is an integer, the degeneracy at the sublevel, where the absolute magnetic quantum numbers are equal, can be solved by taking non-zero values of \(\eta\).
When the nuclear spin \(I\) is larger than \(3/2\), it is possible to uniquely determine the values of \(\eta\) and \(A\) by changing from an isoperimetric sequence of sublevel energy differences \(\eta=0\) to all equal states \(\eta=1\), which is not observable in \(I = 3/2\).
The sign of \(A\) can also be uniquely determined from the shape of the spectrum except for \(I = 3/2\).
Even in the case of \(I = 3/2\), the signs of \(A\) and \(\eta\) can be determined simultaneously from experiments on the basis of the intensity ratio of absorption lines in the spectrum when single crystals are used, so that no posterior probability distribution of uniformly distributed \(\eta\) shown in Figs. \ref{hist_cQ} and \ref{2dhist_cQ} is observed.

\subsubsection{Application of Bayesian inference to a magnetically ordered spectrum}
As shown in Fig. \ref{hist_cQM}, a wide distribution was found for the asymmetric parameter \(\eta\) and the azimuth angle \(\phi\) between the principal axis of the electric field gradient and the internal magnetic field, whereas a good agreement with the true value of the parameter was observed for for the main axis component of the nuclear quadrupole interaction, \(A\).
In addition, a narrow posterior distribution was observed for the angle between the principal axis of the electric field gradient and the internal magnetic field, \(\theta\), which is 0 or \(\pi\).
This is a different behavior from the posterior probability distribution of each parameter in the previous section.
The events that occurred in \(\eta\), \(\phi\), and \(A\) have completely different causes.
The reason for the uniform distribution of \(\phi\) is that the generative model given here is the spectrum of a powder sample.
\(A\), \(\phi\), and \(\eta\) are described in detail as follows.
If the nuclear quadrupole and the nuclear Zeeman interactions act simultaneously, and if the nuclear quadrupole interaction can be treated as a perturbation of the nuclear Zeeman interaction, then the energy of each excited state of \(^{57}\mathrm{Fe}\) \(E_{I,M}\) if each level of the nucleus is described as \(|I,M\rangle\) at \(I = 3/2\) can be expressed as
\begin{align}
	E_{I, M}=-g_{3/2} \mu_{N} B_{hf} M + (-1)^{|M+1/2|} 3A\left(\frac{3\cos^2\theta -1}{2}\right),
	\label{eigen_value_theta_eq}
\end{align}
where \(M\) is the quantum number of nuclear spins in the excited states of \(^{57}\mathrm{Fe}\). 

Assuming now that the main axis component of the nuclear quadrupole interaction, \(A\),  can be treated as a perturbation of the nuclear Zeeman interaction \(g_{3/2} \mu_{N} B_{hf} M\) in the spectrum in Fig. \ref{rowdata_cQM},
we expect to find a relationship similar to that in Fig. \ref{2dhist_cQ} for \(A\) and \(\eta\) so as to satisfy Eq. \eqref{eigen_value_theta_eq}.
For example, \(A = +6\) and \(\theta =- \pi/2\) should also be candidate parameters that reproduce the spectrum.
However, \(A = +6\) is too large to be treated as a perturbation for a true value of the nuclear Zeeman interaction \(B_{hf} = 40\), so it is excluded from the candidates by our analysis method.
Indeed, when the spectrum was calculated using Eqs. \eqref{eigen_value_eq} and \eqref{eigen_value_theta_eq} as \(A = +6\) and \(\theta =- \pi/2\), we confirmed that eight absorption lines were observed instead of six owing to the entanglement of the wave function at each level of \(I = 3/2\).
In other words, this method also verifies that there is no posterior probability distribution at around \(A = +6\) and \(\theta =- \pi/2\), which is predicted under perturbation conditions.
Figure \ref{hist_cQM} shows a very steep posterior probability distribution for the angle between the principal axis of the electric field gradient and the internal magnetic field, \(\theta\), which is \(0\) or \(\pi\).
In this case, the energy of the excited state of \(^{57}\mathrm{Fe}\) \(E_{I,M}\) can be written as follows if each level of the nucleus is described as \(|I,M\rangle\) at \(I = 3/2\).
\begin{align}
	E_{I, M}=\left\{\begin{array}{l}
		\frac{1}{2} g_{3/2} \mu_{N} B_{hf} \pm 3A\left[\left(1+\frac{g \mu_{N} B_{hf}}{3A}\right)^{2}+\frac{\eta^{2}}{3}\right]^{1 / 2} \\
		-\frac{1}{2} g_{3/2} \mu_{N} B_{hf} \pm 3A\left[\left(1-\frac{g \mu_{N} B_{hf}}{3A}\right)^{2}+\frac{\eta^{2}}{3}\right]^{1 / 2}
		\end{array}\right..
	\label{eigen_value_eq}
\end{align}

\begin{figure}
	\centering
	\includegraphics[keepaspectratio,scale=0.7]{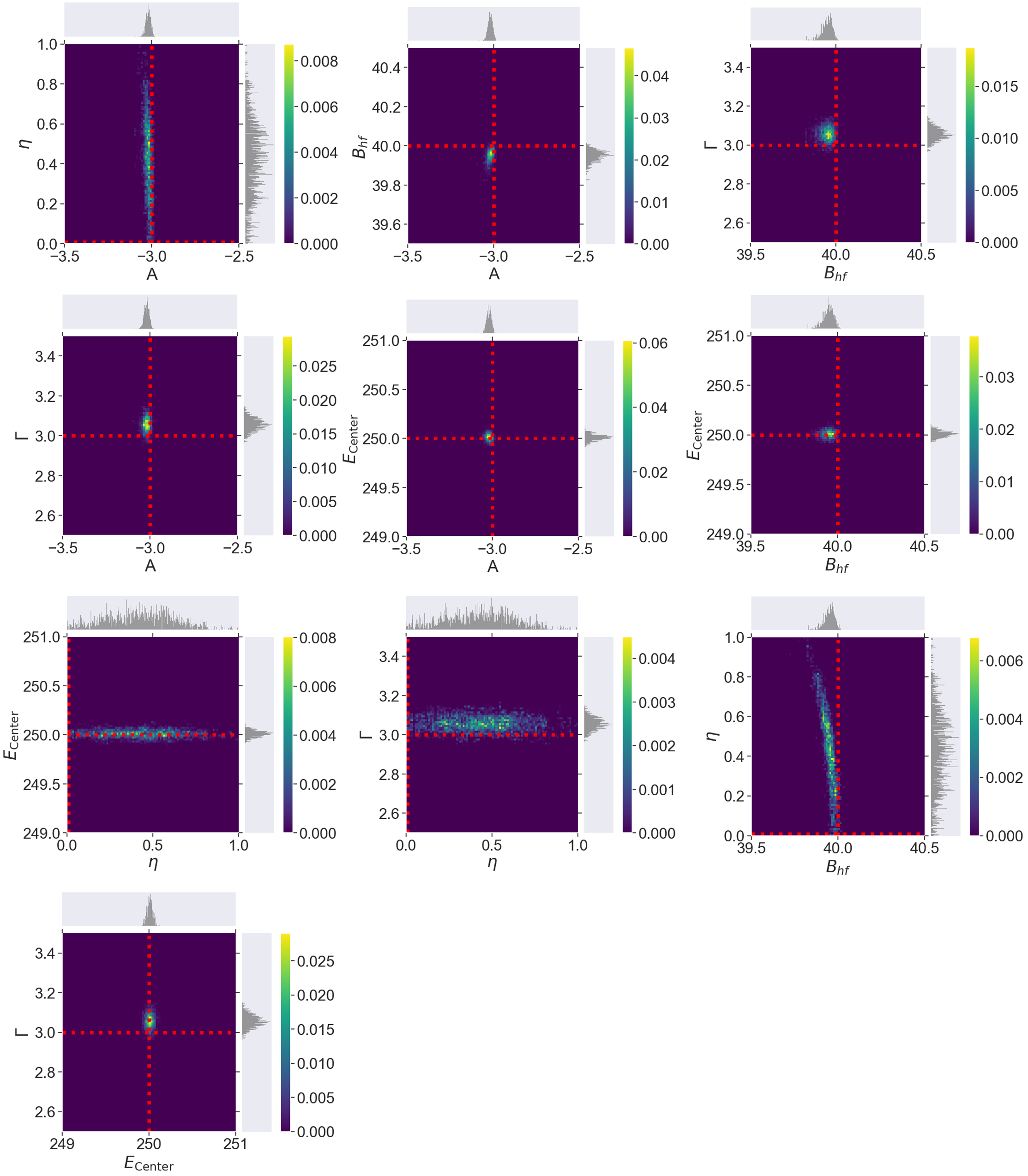}
	\caption{
		(Color online) Two-dimensional histograms of the posterior distributions with the generative model of Hamiltonian \(H_c+H_Q+H_M\) and recognition model of the same Hamiltonian.
		The red dotted lines are the true values of each parameter.}
	\label{2dhist_cQM}
\end{figure}

This energy indicates that the three parameters involved in the nuclear Zeeman and nuclear quadrupole interactions, \(B_{hf}\), \(A\) and \(\eta\), are correlated.
It has been demonstrated that there are correlations in these three parameters in Fig. \ref{2dhist_cQM}.
The narrower range of correlation for \(B_{hf}\) and \(A\) is due to the fact that \(\eta\) is limited to the range of 0 and 1 in Eq. \eqref{eigen_value_eq} and the ratio of \(B_{hf}\) to \(A\) set here is far from 1.
The nonlinear correlations between \(\eta\) and \(B_{hf}\) and between \(\eta\) and \(A\), are the reasons for the asymmetry of the posterior probability distribution shown in Fig. \ref{hist_cQM}.
As discussed in the previous subsection, the posterior probability of \(\eta\) is also uniformly distributed between 0 and 1 in the magnetically ordered case. 
This depends on the peculiar case of hyperfine interactions allowed in the \(I = 3/2\) case. 
Finally, the conditions that provide a monolithic distribution of posterior probability are discussed for \(\eta\) and \(\phi\). \(\eta\) affects transition probabilities observable only in single crystalline samples as well as hyperfine interactions detectable in both single crystalline and polycrystalline samples. 
This means that a sharp distribution of posterior probability in \(\eta\) is expected in single crystalline samples, which is independent of crystallographic and/or electronic properties. On the other hand, the observation of \(\phi\) dependence in spectra requires both the crystallographic orientation of the measured sample and the polarization of the incident X-rays or \(\gamma\)-rays. 
This means that linear or circular polarization, applied magnetic field, or single crystalline samples are required to observe spectra affected by \(\phi\). 
Under such measurement conditions, the distribution of the posterior probability could be given in \(\phi\). 
\subsection{Limits of analyzable noise intensity}
\begin{figure}
	\centering
	\includegraphics[keepaspectratio,scale=0.33]{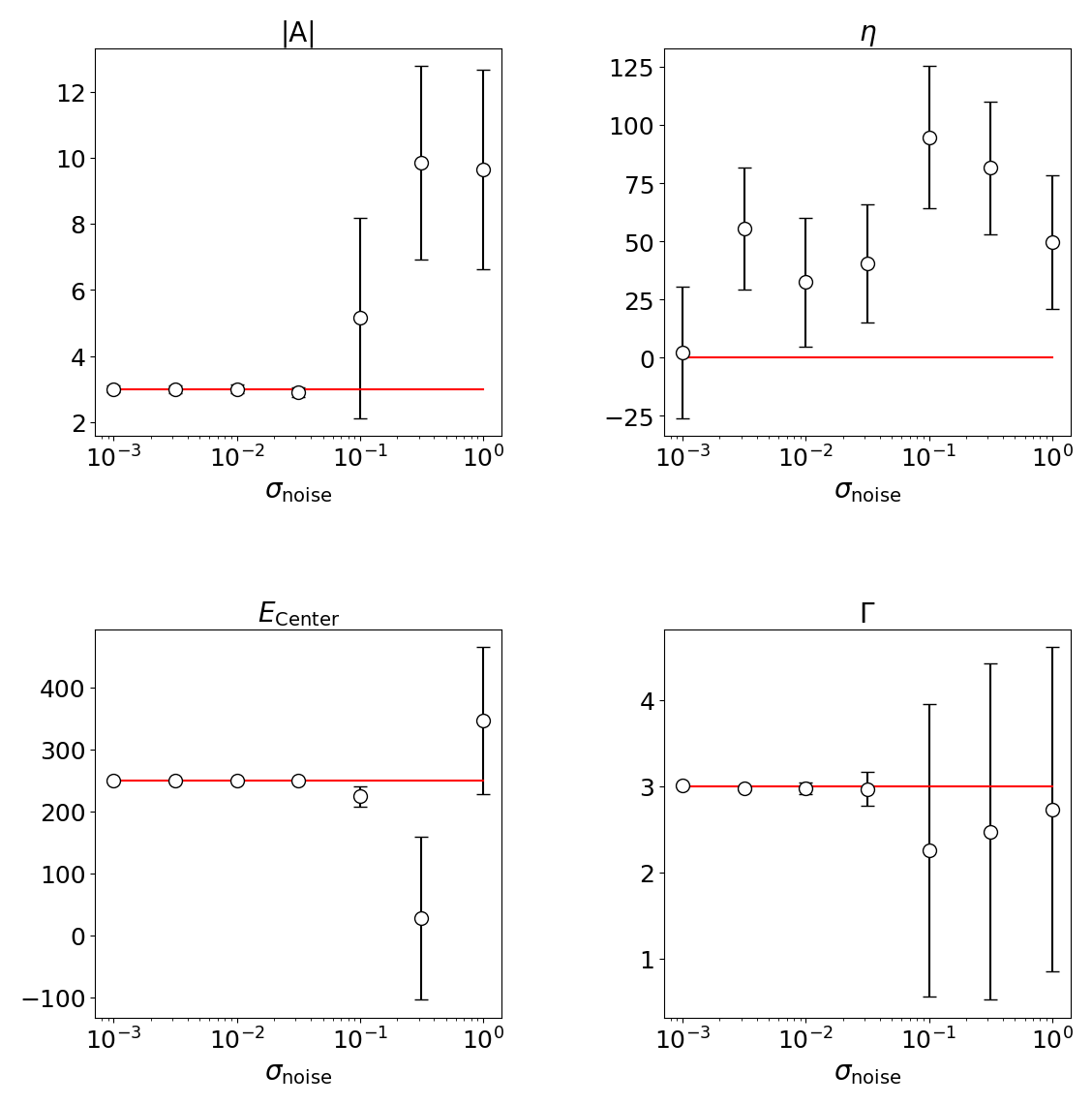}
	\caption{
		    (Color online) Confidence intervals of the estimated values with the generative model of Hamiltonian \(H_c+H_Q\).
			The horizontal axis represents the standard deviation (\(\sigma_{\mathrm{noise}}\)) of the generative model.
			The red lines are the true values of each parameter, the dots are the estimated values, and the error bars represent the confidence intervals.}
	\label{phase_transition_cQ}
\end{figure}

\begin{figure}[htbp]
	\begin{center}
	\subfigure[]{\label{fit_phase1_cQ}%
		\includegraphics[clip, width=0.5\columnwidth]{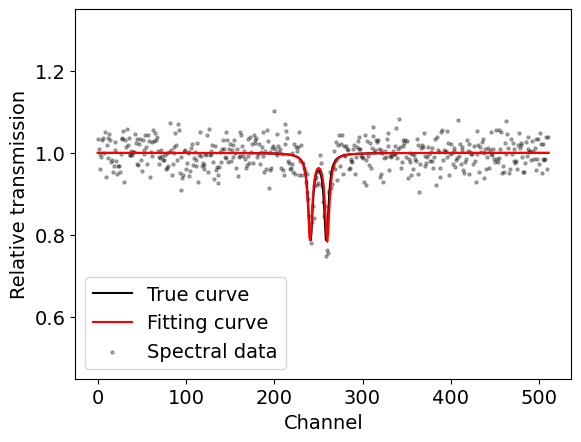}}%
	\subfigure[]{\label{fit_phase2_cQ}%
		\includegraphics[clip, width=0.5\columnwidth]{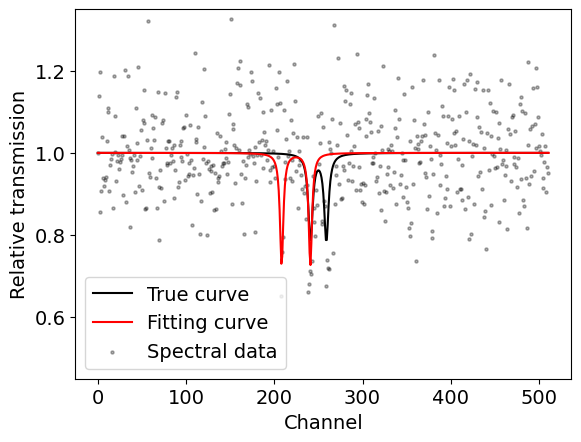}}%
	\caption{
		    (Color online) Results of fitting with generative model of Hamiltonian \(H_c+H_Q\) and recognition model of the same Hamiltonian.
			(a) Spectral data with noise intensity \(\sigma_{\mathrm{noise}}=10^{-1.5}\).
			(b) Spectral data with noise intensity \(\sigma_{\mathrm{noise}}=10^{-1}\).
			The gray dots are spectral data, the red lines are the fitting curves, and the black lines are the true curves of artificial data.}
	\label{fit_phase_transition_cQ} 
	\end{center}
\end{figure}

Figure \ref{phase_transition_cQ} shows the results of the experiment without the nuclear Zeeman interaction, that is, Hamiltonian \(H_c+H_Q\).
Since only absolute values of \(A\) can be obtained from the Hamiltonian, confidence intervals and estimates calculated for \(|A|\) are plotted.
From Fig. \ref{phase_transition_cQ}, we can see that in the absence of the nuclear Zeeman interaction, the confidence interval rapidly widens around the noise intensity \(\sigma_{\mathrm{noise}}=10^{-1.5}\).
Here, the fitting results at around the noise intensity \(\sigma_{\mathrm{noise}}=10^{-1.5}\) are shown in Fig. \ref{fit_phase_transition_cQ} where fitting is possible for the noise intensity \(\sigma_{\mathrm{noise}}=10^{-1.5}\), whereas fitting is not possible for the noise intensity \(\sigma_{\mathrm{noise}}=10^{-1}\).
From the above, it was found that the data without the nuclear Zeeman interaction can be estimated up to the noise intensity \(\sigma_{\mathrm{noise}}=10^{-1.5}\).

\begin{figure}
	\centering
	\includegraphics[keepaspectratio,scale=0.36]{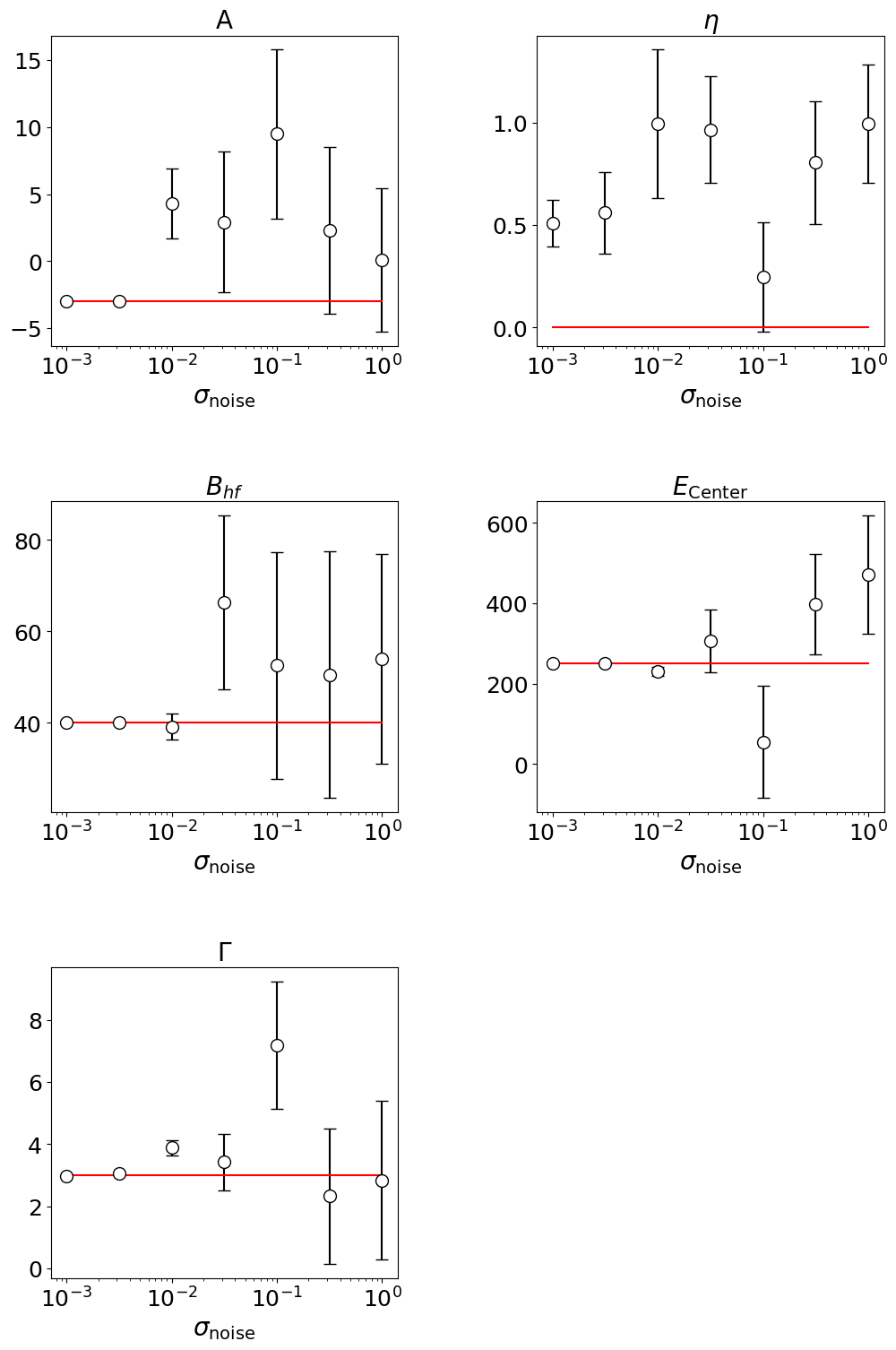}
	\caption{
		    (Color online) Confidence intervals of the estimated values with generative model of Hamiltonian \(H_c+H_Q+H_M\).
			The horizontal axis represents the standard deviation (\(\sigma_{\mathrm{noise}}\)) of the generative model.
			The red lines are the true values of each parameter, the dots are the estimated values, and the error bars are the confidence intervals.}
	\label{phase_transition_cQM}
\end{figure}

\begin{figure}[htbp]
	\begin{center}		
	\subfigure[]{\label{fit_phase1_cQM}%
		\includegraphics[clip, width=0.33\columnwidth]{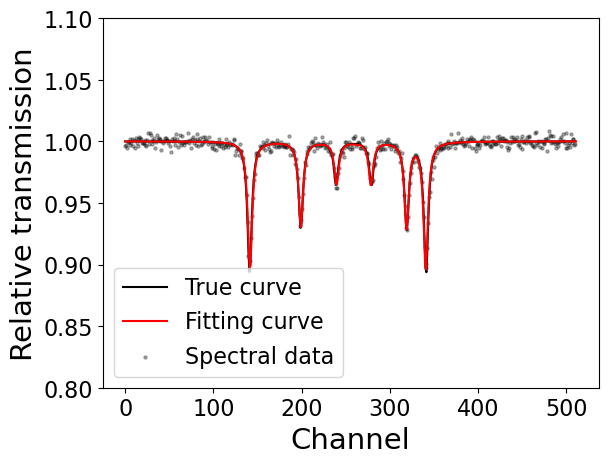}}%
	\subfigure[]{\label{fit_phase2_cQM}%
		\includegraphics[clip, width=0.33\columnwidth]{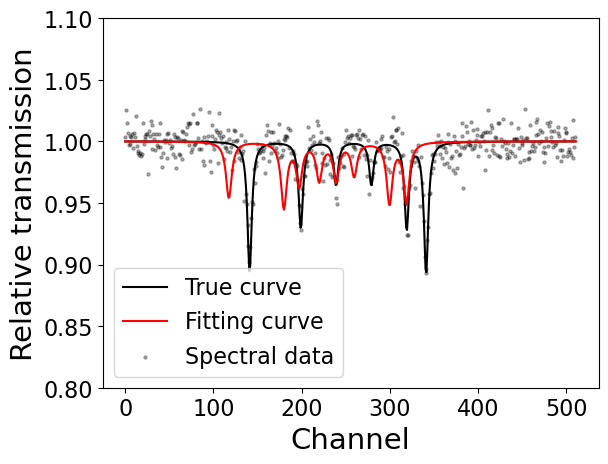}}%
	\subfigure[]{\label{fit_phase3_cQM}%
		\includegraphics[clip, width=0.33\columnwidth]{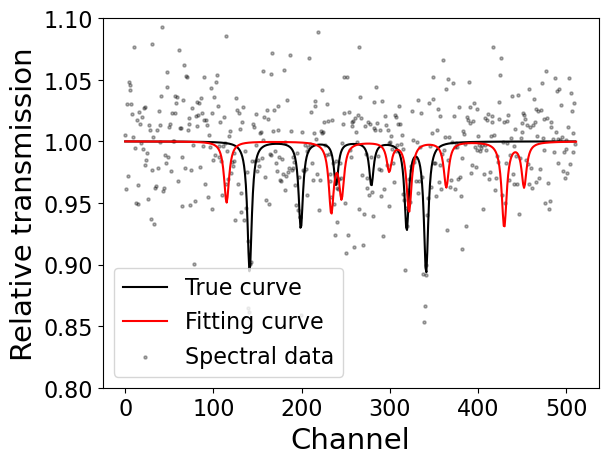}}%
	\caption{
		    (Color online) Results of fitting with the generative model of Hamiltonian \(H_c+H_Q+H_M\) and recognition model of the same Hamiltonian.
			Spectral data with noise intensities (a) \(\sigma_{\mathrm{noise}}=10^{-2.5}\),
			(b) \(\sigma_{\mathrm{noise}}=10^{-2}\), and
			(c) \(\sigma_{\mathrm{noise}}=10^{-1.5}\).}
	\label{fit_phase_transition_cQM} 
	\end{center}
\end{figure}

Next, Fig. \ref{phase_transition_cQM} shows the results for the experiment with the nuclear Zeeman interaction, that is, Hamiltonian \(H_c+H_Q+H_M\).
From Fig. \ref{phase_transition_cQM}, the confidence interval rapidly widens around the noise intensity \(\sigma_{\mathrm{noise}}=10^{-2.5}\).
Here, the fitting results around the noise intensity \(\sigma_{\mathrm{noise}}=10^{-2.5}\) are shown in Fig. \ref{fit_phase_transition_cQM} where fitting is possible for the noise intensity \(\sigma_{\mathrm{noise}}=10^{-2.5}\), but not for the noise intensity \(\sigma_{\mathrm{noise}}=10^{-2}\).
From the above, it was found that the data without the nuclear Zeeman interaction can be estimated up to a noise strength of \(\sigma_{\mathrm{noise}}=10^{-2.5}\).

The results of this experiment show that for the spectrum of a single site, the standard deviation of the noise is less than \(\sigma_{\mathrm{noise}}=10^{-1.5}\) when there is no nuclear Zeeman interaction.
In the case of the nuclear Zeeman interaction, the spectral data are less than or equal to \(\sigma_{\mathrm{noise}}=10^{-2.5}\).
The above results enable us to evaluate the accuracy to which data can be analyzed when conducting Mössbauer spectroscopy experiments or when analyzing measurement data.

\section{Conclusion}
We have applied Bayesian inference to Hamiltonian selection on hyperfine interactions for analyses of \(^{57}\mathrm{Fe}\) Mössbauer spectra without prior information. 
The results obtained in this work exhibit the applicability of spectral analyses in two cases, paramagnetic and magnetically ordered spectra, by Hamiltonian selection based on Bayesian inference without prior information. 
Using the present results of the posterior distribution obtained by Bayesian inference, we have successfully calculated the posterior standard distribution, whose reliability is difficult to evaluate in conventional least-square analyses. 

Taking advantage of the characteristics of the generative model used, we evaluated confidence intervals at various noise intensities of the model. 
The changes in confidence intervals and MAP estimates of noise intensity indicate the accuracy required when using this analysis method on actual experimental data. 

This work demonstrates that Bayesian inference without prior information from other experiments such as crystallography and magnetism enabled the successfull extraction of hyperfine interactions at the probe nuclear sites from the \(^{57}\mathrm{Fe}\) Mössbauer spectra. 
One of the reasons for the success is that the description of the hyperfine interaction at a single site is very limited in \(^{57}\mathrm{Fe}\) Mössbauer spectroscopy. 
This method is applicable to other isotopes as well as \(^{57}\mathrm{Fe}\) nuclei, as reported in this paper. 
When the observable regions of the isomer shifts in the probe nuclei are provided as prior information, more accurate analyses based on Bayesian inference is possible. 
Similarly, this work is also applicable to the analyses of spectra obtained by NMR/NQR and PAC spectroscopies, because such spectra are described by hyperfine interactions at the probe nuclear sites. 


\bibliographystyle{jpsj}
\bibliography{bibtex_jpsj}

\end{document}